<?xml version="1.0" encoding="UTF-8"?>
<!DOCTYPE html PUBLIC "-//W3C//DTD XHTML 1.0 Transitional//EN" "http://www.w3.org/TR/xhtml1/DTD/xhtml1-transitional.dtd">
<html xmlns="http://www.w3.org/1999/xhtml" lang="en">
<head>
  <link rel="shortcut icon" href="/favicon.ico" type="image/x-icon" />
<link rel="stylesheet" type="text/css" media="screen" href="//static.arxiv.org/css/arXiv.css?v=20170424" />
  <!-- Piwik -->
  <script type="text/javascript">
    var _paq = _paq || [];
    _paq.push(["setDomains", ["*.arxiv.org"]]);
    _paq.push(['trackPageView']);
    _paq.push(['enableLinkTracking']);
    (function() {
      var u = "//webanalytics.library.cornell.edu/";
      _paq.push(['setTrackerUrl', u + 'piwik.php']);
      _paq.push(['setSiteId', 538]);
      var d = document,
        g = d.createElement('script'),
        s = d.getElementsByTagName('script')[0];
      g.type = 'text/javascript';
      g.async = true;
      g.defer = true;
      g.src = u + 'piwik.js';
      s.parentNode.insertBefore(g, s);
    })();
  </script>
  <!-- End Piwik Code -->
  <script type="text/javascript" src="https://culibrary.atlassian.net/s/d41d8cd98f00b204e9800998ecf8427e-T/-w6uc8u/b/16/a44af77267a987a660377e5c46e0fb64/_/download/batch/com.atlassian.jira.collector.plugin.jira-issue-collector-plugin:issuecollector/com.atlassian.jira.collector.plugin.jira-issue-collector-plugin:issuecollector.js?locale=en-US&collectorId=7a8da419"></script>
<script type="text/javascript">window.ATL_JQ_PAGE_PROPS =  {
  "triggerFunction": function(showCollectorDialog) {
    //Requires that jQuery is available!
    jQuery("#feedback-button").click(function(e) {
      e.preventDefault();
      showCollectorDialog();
    });
  },
  fieldValues: {
    "components": ["15700"],  // Browse component.
    "versions": ["14132"],  // Release browse-0.1
    "customfield_11401": window.location.href
  }
  };
</script><link rel="stylesheet" media="screen" type="text/css" href="/bibex/bibex.css?20181010"/>
<script type="text/x-mathjax-config">
  MathJax.Hub.Config({
    messageStyle: "none",
    extensions: ["tex2jax.js"],
    jax: ["input/TeX", "output/HTML-CSS"],
    tex2jax: {
      inlineMath: [ ['$','$'], ["\\(","\\)"] ],
      displayMath: [ ['$$','$$'], ["\\[","\\]"] ],
      processEscapes: true,
      ignoreClass: '.*',
      processClass: 'mathjax.*'
    },
    TeX: {
        extensions: ["AMSmath.js", "AMSsymbols.js", "noErrors.js"],
        noErrors: {
          inlineDelimiters: ["$","$"],
          multiLine: false,
          style: {
            "font-size": "normal",
            "border": ""
          }
        }
    },
    "HTML-CSS": { availableFonts: ["TeX"] }
  });
</script>
<script src='//static.arxiv.org/MathJax-2.7.3/MathJax.js'></script>
<meta name="citation_title" content="Proceedings of the fourth &#34;international Traveling Workshop on Interactions between low-complexity data models and Sensing Techniques&#34; (iTWIST&#39;18)"/>
  <meta name="citation_author" content="Anthoine, Sandrine"/>
  <meta name="citation_author" content="Boursier, Yannick"/>
  <meta name="citation_author" content="Jacques, Laurent"/>
  <meta name="citation_date" content="2018/12/03"/>
  <meta name="citation_online_date" content="2018/12/03"/>
  <meta name="citation_pdf_url" content="https://arxiv.org/pdf/1812.00648"/>
  <meta name="citation_arxiv_id" content="1812.00648"/>
<title>iTWIST'18 Proceedings</title>
</head>
<body>
<center>
<h1>iTWIST: international Traveling Workshop on Interactions<br /> between low-complexity data models and Sensing Techniques</h1>
<h2>iTWIST'18 Proceedings</h2>
<a href="https://sites.google.com/view/itwist18">Workshop website</a><br />
<b>Editors:</b> Sandrine Anthoine, Yannick Boursier, Laurent Jacques<br />
<b>Scientific committee:</b> <a href="https://sites.google.com/view/itwist18/committees">(see here)</a><br /><br />
﻿CIRM, Marseille, France<br />
21 - 23 November​, 2018<br />
</center>

<p>The iTWIST workshop series aim at fostering collaboration between international scientific teams for developing new theories, applications and generalizations of low-complexity models. These events emphasize dissemination of ideas through both specific oral and poster presentations, as well as free discussions. For this fourth edition, iTWIST'18 gathered 74 international participants and featured 7 invited talks, 16 oral presentations, and 21 posters.</p>

<p>From this edition, the scientific committee has decided that the iTWIST'18 proceedings will adopt the <a href="https://www.episciences.org/">episcience.org</a> philosophy, combined with <a href="http://arxiv.org">arXiv.org<a>: in a nutshell, <i>"the proceedings are equivalent to an overlay page, built above <a href="http://arxiv.org">arXiv.org<a>; they add value to these archives by attaching a scientific caution to the validated papers."</i></p>

<p>For iTWIST'18, this means that <b>all papers listed below have been thoroughly evaluated and approved by two independent reviewers, and authors have revised their work according to the comments provided by these reviewers.</b></p>

<hr>
<!-- proceeding entry -->
<h3>Proceedings of iTWIST'18, Paper-ID: 1, Marseille, France, November, 21-23, 2018. </h3>
<dl>
<dd>
<div class="meta">
<div class="list-title mathjax"> <span class="descriptor">Title: "A hierarchical Bayesian perspective on majorization-minimization for non-convex sparse regression"</span> </div>
<div class="list-authors"><span class="descriptor">Authors: Y. Bekhti, F. Lucka, J. Salmon and A. Gramfort</span> </div>
</div>
</dd>
</dl>
<i>Remark: This iTWIST'18 2-page paper is related to this longer article by
the same authors: <a href="https://arxiv.org/abs/1710.08747">arXiv:1710.08747</a></i>

<hr>
<!-- proceeding entry -->
<h3>Proceedings of iTWIST'18, Paper-ID: 2, Marseille, France, November, 21-23, 2018. </h3>
<!-- <dl> -->
<!-- <dd> -->
<!-- <div class="meta"> -->
<!-- <div class="list-title mathjax"> <span class="descriptor">Title: "A scalable estimator of sets of integral operators"</span> </div> -->
<!-- <div class="list-authors"><span class="descriptor">Authors: V. Debarnot, P. Escande and P. Weiss</span> </div> -->
<!-- </div> -->
<!-- </dd> -->
<!-- </dl> -->
LIST:arXiv:1811.12192

<hr>
<!-- proceeding entry -->
<h3>Proceedings of iTWIST'18, Paper-ID: 3, Marseille, France, November, 21-23, 2018.</h3>
<dl>
<dd>
<div class="meta">
<div class="list-title mathjax"> <span class="descriptor">Title: "Joint-sparse modeling for audio inpainting"</span> </div>
<div class="list-authors"><span class="descriptor">Authors: I. Toumi and V. Emiya</span> </div>
</div>
</dd>
</dl>

<hr>
<!-- proceeding entry -->
<h3>Proceedings of iTWIST'18, Paper-ID: 4, Marseille, France, November, 21-23, 2018. </h3>
<!-- <dl> -->
<!-- <dd> -->
<!-- <div class="meta"> -->
<!-- <div class="list-title mathjax"> <span class="descriptor">Title: "Fast Iterative Shrinkage for Signal Declipping and Dequantization"</span> </div> -->
<!-- <div class="list-authors"><span class="descriptor">Authors: L. Rencker, F. Bach, W. Wang and M. Plumbley</span> </div> -->
<!-- </div> -->
<!-- </dd> -->
<!-- </dl> -->
LIST:arXiv:1812.01540

<hr>
<h3>Proceedings of iTWIST'18, Paper-ID: 5, Marseille, France, November, 21-23, 2018.</h3>
<!-- S. Guérit, S. Sivankutty, C. Scotté, J. A. Lee, H. Rigneault and L. Jacques, ``Compressive Sampling Approach for Image Acquisition with Lensless Endoscope'' -->
LIST:arXiv:1810.12286

<hr>
<h3>Proceedings of iTWIST'18, Paper-ID: 6, Marseille, France, November, 21-23, 2018.</h3>
<!-- proceeding entry -->
<dl>
<dd>
<div class="meta">
<div class="list-title mathjax"> <span class="descriptor">Title: "A Dual Certificates Analysis of Compressive Off-the-Grid Recovery"</span> </div>
<div class="list-authors"><span class="descriptor">Authors: C. Poon, N. Keriven and G. Peyré</span> </div>
</div>
</dd>
</dl>
<i>Remark: This iTWIST'18 2-page paper is related to these longer articles by
the same authors: <a
href="https://arxiv.org/abs/1810.03340">arXiv:1810.03340</a> and <a
href="https://arxiv.org/abs/1802.08464">arXiv:1802.08464</a>
</i>

<hr>
<h3>Proceedings of iTWIST'18, Paper-ID: 7, Marseille, France, November, 21-23, 2018.</h3>
<!-- A. Moshtaghpour, L. Jacques and J. M. Bioucas Dias, ``Compressive Hyperspectral Imaging: Fourier Transform Interferometry meets Single Pixel Camera'' -->
LIST:arXiv:1809.00950

<hr>
<h3>Proceedings of iTWIST'18, Paper-ID: 8, Marseille, France, November, 21-23, 2018.</h3>
<!-- proceeding entry -->
<!-- <dl> -->
<!-- <dd> -->
<!-- <div class="meta"> -->
<!-- <div class="list-title mathjax"> <span class="descriptor">Title: "Compressive Classification (Machine Learning without learning)"</span> </div> -->
<!-- <div class="list-authors"><span class="descriptor">Authors: V. Schellekens and L. Jacques</span> </div> -->
<!-- </div> -->
<!-- </dd> -->
<!-- </dl> -->
LIST:arXiv:1812.01410

<hr>
<h3>Proceedings of iTWIST'18, Paper-ID: 9, Marseille, France, November, 21-23, 2018.</h3>
<!-- proceeding entry -->
<!-- <dl> -->
<!-- <dd> -->
<!-- <div class="meta"> -->
<!-- <div class="list-title mathjax"> <span class="descriptor">Title: "Sparse component separation from Poisson measurements"</span> </div> -->
<!-- <div class="list-authors"><span class="descriptor">Authors: I. El Hamzaoui and J. Bobin</span> </div> -->
<!-- </div> -->
<!-- </dd> -->
<!-- </dl> -->
LIST:arXiv:1812.04370

<hr>
<h3>Proceedings of iTWIST'18, Paper-ID: 10, Marseille, France, November, 21-23, 2018.</h3>
<!-- proceeding entry -->
<!-- <dl> -->
<!-- <dd> -->
<!-- <div class="meta"> -->
<!-- <div class="list-title mathjax"> <span class="descriptor">Title: "BSGD-TV:A parallel algorithm solving total variation constrained image reconstruction problems"</span> </div> -->
<!-- <div class="list-authors"><span class="descriptor">Authors: Y. Gao and T. Blumensath</span> </div> -->
<!-- </div> -->
<!-- </dd> -->
<!-- </dl> -->
LIST:arXiv:1812.01307

<hr>
<h3>Proceedings of iTWIST'18, Paper-ID: 11, Marseille, France, November, 21-23, 2018.</h3>
<!-- proceeding entry -->
<!-- <dl> -->
<!-- <dd> -->
<!-- <div class="meta"> -->
<!-- <div class="list-title mathjax"> <span class="descriptor">Title: "Heuristics for Efficient Sparse Blind Source Separation"</span> </div> -->
<!-- <div class="list-authors"><span class="descriptor">Authors: C. Kervazo, J. Bobin and C. Chenot</span> </div> -->
<!-- </div> -->
<!-- </dd> -->
<!-- </dl> -->
LIST:arXiv:1812.06737

<hr>
<h3>Proceedings of iTWIST'18, Paper-ID: 12, Marseille, France, November, 21-23, 2018.</h3>
<!-- proceeding entry -->
<!-- <dl> -->
<!-- <dd> -->
<!-- <div class="meta"> -->
<!-- <div class="list-title mathjax"> <span class="descriptor">Title: "Is the 1-norm the best convex sparse regularization?"</span> </div> -->
<!-- <div class="list-authors"><span class="descriptor">Authors: Y. Traonmilin, S. Vaiter and R. Gribonval</span> </div> -->
<!-- </div> -->
<!-- </dd> -->
<!-- </dl> -->
LIST:arXiv:1806.08690

<hr>
<h3>Proceedings of iTWIST'18, Paper-ID: 13, Marseille, France, November, 21-23, 2018.</h3>
<!-- proceeding entry -->
<!-- <dl> -->
<!-- <dd> -->
<!-- <div class="meta"> -->
<!-- <div class="list-title mathjax"> <span class="descriptor">Title: "Joint nonstationary blind source separation and spectral analysis"</span> </div> -->
<!-- <div class="list-authors"><span class="descriptor">Authors: A. Meynard</span> </div> -->
<!-- </div> -->
<!-- </dd> -->
<!-- </dl> -->
LIST:arXiv:1812.01399

<hr>
<h3>Proceedings of iTWIST'18, Paper-ID: 14, Marseille, France, November, 21-23, 2018.</h3>
<!-- proceeding entry -->
<!-- <dl> -->
<!-- <dd> -->
<!-- <div class="meta"> -->
<!-- <div class="list-title mathjax"> <span class="descriptor">Title: "Efficient atom selection strategy for iterative sparse approximations"</span> </div> -->
<!-- <div class="list-authors"><span class="descriptor">Authors: C. Dorffer, A. Drémeau and C. Herzet</span> </div> -->
<!-- </div> -->
<!-- </dd> -->
<!-- </dl> -->
LIST:arXiv:1812.01932

<hr>
<h3>Proceedings of iTWIST'18, Paper-ID: 15, Marseille, France, November, 21-23, 2018.</h3>
<!-- proceeding entry -->
<!-- <dl> -->
<!-- <dd> -->
<!-- <div class="meta"> -->
<!-- <div class="list-title mathjax"> <span class="descriptor">Title: "Data-driven cortical clustering to provide a family of plausible solutions to M/EEG inverse problem"</span> </div> -->
<!-- <div class="list-authors"><span class="descriptor">Authors: K. Maksymenko, M. Clerc and T. Papadopoulo</span> </div> -->
<!-- </div> -->
<!-- </dd> -->
<!-- </dl> -->
LIST:arXiv:1812.04110

<hr>
<h3>Proceedings of iTWIST'18, Paper-ID: 16, Marseille, France, November, 21-23, 2018.</h3>
<!-- proceeding entry -->
<!-- <dl> -->
<!-- <dd> -->
<!-- <div class="meta"> -->
<!-- <div class="list-title mathjax"> <span class="descriptor">Title: "Performance Analysis of Approximate Message Passing for Distributed Compressed Sensing"</span> </div> -->
<!-- <div class="list-authors"><span class="descriptor">Authors: G. Hannak, A. Perelli, G. Matz, M. Davies and N. Goertz</span> </div> -->
<!-- </div> -->
<!-- </dd> -->
<!-- </dl> -->
LIST:arXiv:1712.04893

<hr>
<h3>Proceedings of iTWIST'18, Paper-ID: 17, Marseille, France, November, 21-23, 2018.</h3>
<!-- proceeding entry -->
<!-- <dl> -->
<!-- <dd> -->
<!-- <div class="meta"> -->
<!-- <div class="list-title mathjax"> <span class="descriptor">Title: "From biological vision to unsupervised hierarchical sparse coding"</span> </div> -->
<!-- <div class="list-authors"><span class="descriptor">Authors: V. Boutin, A. Franciosini, F. Ruffier and L. Perrinet</span> </div> -->
<!-- </div> -->
<!-- </dd> -->
<!-- </dl> -->
LIST:arXiv:1812.01335

<hr>
<h3>Proceedings of iTWIST'18, Paper-ID: 18, Marseille, France, November, 21-23, 2018.</h3>
<!-- <\!-- to remove once submitted -\-> -->
<!-- <dl> -->
<!-- <dd> -->
<!-- <div class="meta"> -->
<!-- <div class="list-title mathjax"> <span class="descriptor">Title: "Denoising and Completion of Structured Low-Rank Matrices via Iteratively Reweighted Least Squares"</span> </div> -->
<!-- <div class="list-authors"><span class="descriptor">Authors: C.Kümmerle and C. Mayrink Verdun</span> </div> -->
<!-- </div> -->
<!-- </dd> -->
<!-- </dl> -->
LIST:arXiv:1811.07472

<hr>
<h3>Proceedings of iTWIST'18, Paper-ID: 19, Marseille, France, November, 21-23, 2018.</h3>
<!-- proceeding entry -->
<!-- <dl> -->
<!-- <dd> -->
<!-- <div class="meta"> -->
<!-- <div class="list-title mathjax"> <span class="descriptor">Title: "Blended smoothing splines on Riemannian manifolds"</span> </div> -->
<!-- <div class="list-authors"><span class="descriptor">Authors: P.-Y. Gousenbourger, E. Massart and P.-A. Absil</span> </div> -->
<!-- </div> -->
<!-- </dd> -->
<!-- </dl> -->
LIST:arXiv:1812.04420

<hr>
<h3>Proceedings of iTWIST'18, Paper-ID: 20, Marseille, France, November, 21-23, 2018.</h3>
<!-- proceeding entry -->
<!-- <dl> -->
<!-- <dd> -->
<!-- <div class="meta"> -->
<!-- <div class="list-title mathjax"> <span class="descriptor">Title: "From Adaptive Kernel Density Estimation to Sparse Mixture Models"</span> </div> -->
<!-- <div class="list-authors"><span class="descriptor">Authors: C. Schretter, J. Sun and P. Schelkens</span> </div> -->
<!-- </div> -->
<!-- </dd> -->
<!-- </dl> -->
LIST:arXiv:1812.04397

<hr>
<h3>Proceedings of iTWIST'18, Paper-ID: 21, Marseille, France, November, 21-23, 2018.</h3>
<!-- proceeding entry -->
<!-- <dl> -->
<!-- <dd> -->
<!-- <div class="meta"> -->
<!-- <div class="list-title mathjax"> <span class="descriptor">Title: "A Low-Rank and Joint-Sparse Model for Ultrasound Signal Reconstruction"</span> </div> -->
<!-- <div class="list-authors"><span class="descriptor">Authors: M. Zhang, I. Markovsky, C. Schretter and J. D'Hooge</span> </div> -->
<!-- </div> -->
<!-- </dd> -->
<!-- </dl> -->
LIST:arXiv:1812.04843

<hr>
<h3>Proceedings of iTWIST'18, Paper-ID: 22, Marseille, France, November, 21-23, 2018.</h3>
<!-- proceeding entry -->
<dl>
<dd>
<div class="meta">
<div class="list-title mathjax"> <span class="descriptor">Title: "Multi-scale Decomposition of Transformation (MUSCADET)"</span> </div>
<div class="list-authors"><span class="descriptor">Authors: P. Escande and M. Maggioni</span> </div>
</div>
</dd>
</dl>

<hr>
<h3>Proceedings of iTWIST'18, Paper-ID: 23, Marseille, France, November, 21-23, 2018.</h3>
<!-- proceeding entry -->
<!-- <dl> -->
<!-- <dd> -->
<!-- <div class="meta"> -->
<!-- <div class="list-title mathjax"> <span class="descriptor">Title: "Reference-less algorithm for circumstellar disks imaging"</span> </div> -->
<!-- <div class="list-authors"><span class="descriptor">Authors: B. Pairet, L. Jacques and F. Cantalloube</span> </div> -->
<!-- </div> -->
<!-- </dd> -->
<!-- </dl> -->
LIST:arXiv:1812.01333

<hr>
<h3>Proceedings of iTWIST'18, Paper-ID: 24, Marseille, France, November, 21-23, 2018.</h3>
<!-- proceeding entry -->
<!-- <dl> -->
<!-- <dd> -->
<!-- <div class="meta"> -->
<!-- <div class="list-title mathjax"> <span class="descriptor">Title: "Generalised Approximate Message Passing for Non-I.I.D. Sparse Signals"</span> </div> -->
<!-- <div class="list-authors"><span class="descriptor">Authors: C. Schou Oxvig and T. Arildsen</span> </div> -->
<!-- </div> -->
<!-- </dd> -->
<!-- </dl> -->
LIST:arXiv:1812.00909

<hr>
<h3>Proceedings of iTWIST'18, Paper-ID: 25, Marseille, France, November, 21-23, 2018.</h3>
<!-- proceeding entry -->
<!-- <dl> -->
<!-- <dd> -->
<!-- <div class="meta"> -->
<!-- <div class="list-title mathjax"> <span class="descriptor">Title: "Frank-Wolfe Algorithm for the m-EXACT-SPARSE Problem"</span> </div> -->
<!-- <div class="list-authors"><span class="descriptor">Authors: F. Cherfaoui, V. Emiya, L. Ralaivola and S. Anthoine</span> </div> -->
<!-- </div> -->
<!-- </dd> -->
<!-- </dl> -->
LIST:arXiv:1812.07201

<hr>
<h3>Proceedings of iTWIST'18, Paper-ID: 26, Marseille, France, November, 21-23, 2018.</h3>
<!-- proceeding entry -->
<!-- <dl> -->
<!-- <dd> -->
<!-- <div class="meta"> -->
<!-- <div class="list-title mathjax"> <span class="descriptor">Title: "Learning Discrimiative Representation with Signed Laplacian Restricted Boltzmann Machine"</span> </div> -->
<!-- <div class="list-authors"><span class="descriptor">Authors: D. Chen, J. Lv and M. Davies</span> </div> -->
<!-- </div> -->
<!-- </dd> -->
<!-- </dl> -->
LIST:arXiv:1808.09389

<hr>
<h3>Proceedings of iTWIST'18, Paper-ID: 27, Marseille, France, November, 21-23, 2018.</h3>
<!-- proceeding entry -->
<!-- <dl> -->
<!-- <dd> -->
<!-- <div class="meta"> -->
<!-- <div class="list-title mathjax"> <span class="descriptor">Title: "Matrix Factorization via Deep Learning"</span> </div> -->
<!-- <div class="list-authors"><span class="descriptor">Authors: D. Minh Nguyen, E. Tsiligianni and N. Deligiannis</span> </div> -->
<!-- </div> -->
<!-- </dd> -->
<!-- </dl> -->
LIST:arXiv:1812.01478

<hr>
<h3>Proceedings of iTWIST'18, Paper-ID: 28, Marseille, France, November, 21-23, 2018.</h3>
<!-- proceeding entry -->
<!-- <dl> -->
<!-- <dd> -->
<!-- <div class="meta"> -->
<!-- <div class="list-title mathjax"> <span class="descriptor">Title: "Faster-than-fast NMF using random projections and Nesterov iterations"</span> </div> -->
<!-- <div class="list-authors"><span class="descriptor">Authors: F. Yahaya, M. Puigt, G. Delmaire and G. Roussel</span> </div> -->
<!-- </div> -->
<!-- </dd> -->
<!-- </dl> -->
LIST:arXiv:1812.04315

<hr>
<h3>Proceedings of iTWIST'18, Paper-ID: 29, Marseille, France, November, 21-23, 2018.</h3>
<!-- proceeding entry -->
<!-- <dl> -->
<!-- <dd> -->
<!-- <div class="meta"> -->
<!-- <div class="list-title mathjax"> <span class="descriptor">Title: "An extreme bit-rate reduction scheme for 2D radar localization"</span> </div> -->
<!-- <div class="list-authors"><span class="descriptor">Authors: T. Feuillen, L. Vandendorpe and L. Jacques</span> </div> -->
<!-- </div> -->
<!-- </dd> -->
<!-- </dl> -->
LIST:arXiv:1812.05359

<hr>
<h3>Proceedings of iTWIST'18, Paper-ID: 30, Marseille, France, November, 21-23, 2018.</h3>
<!-- proceeding entry -->
<!-- <dl> -->
<!-- <dd> -->
<!-- <div class="meta"> -->
<!-- <div class="list-title mathjax"> <span class="descriptor">Title: "Convex Regularization and Representer Theorems"</span> </div> -->
<!-- <div class="list-authors"><span class="descriptor">Authors: C. Boyer, A. Chambolle, Y. De Castro, V. Duval, F. de Gournay and P. Weiss</span> </div> -->
<!-- </div> -->
<!-- </dd> -->
<!-- </dl> -->
LIST:arXiv:1812.04355

<hr>
<h3>Proceedings of iTWIST'18, Paper-ID: 31, Marseille, France, November, 21-23, 2018.</h3>
<!-- proceeding entry -->
<!-- <dl> -->
<!-- <dd> -->
<!-- <div class="meta"> -->
<!-- <div class="list-title mathjax"> <span class="descriptor">Title: "Stability of Scattering Decoder for Nonlinear Diffractive Imaging"</span> </div> -->
<!-- <div class="list-authors"><span class="descriptor">Authors: Y. Sun and U. Kamilov</span> </div> -->
<!-- </div> -->
<!-- </dd> -->
<!-- </dl> -->
LIST:arXiv:1806.08015

<hr>
<h3>Proceedings of iTWIST'18, Paper-ID: 32, Marseille, France, November, 21-23, 2018.</h3>
<!-- proceeding entry -->
<dl>
<dd>
<div class="meta">
<div class="list-title mathjax"> <span class="descriptor">Title: "A simple approach to hierarchical classification"</span> </div>
<div class="list-authors"><span class="descriptor">Authors: D. Molitor and D. Needell</span> </div>
</div>
</dd>
</dl>
<i>Remark: This iTWIST'18 2-page paper is related to this longer article by
the same authors: <a href="https://arxiv.org/abs/1807.08825">arXiv:1807.08825</a></i>

<hr>
<h3>Proceedings of iTWIST'18, Paper-ID: 33, Marseille, France, November, 21-23, 2018.</h3>
<!-- proceeding entry -->
<!-- <dl> -->
<!-- <dd> -->
<!-- <div class="meta"> -->
<!-- <div class="list-title mathjax"> <span class="descriptor">Title: "Semi-supervised dual graph regularized dictionary learning"</span> </div> -->
<!-- <div class="list-authors"><span class="descriptor">Authors: K.-H. Tran, F.-M. Ngole-Mboula and J.-L. Starck</span> </div> -->
<!-- </div> -->
<!-- </dd> -->
<!-- </dl> -->
LIST:arXiv:1812.04456

<hr>
<h3>Proceedings of iTWIST'18, Paper-ID: 34, Marseille, France, November, 21-23, 2018.</h3>
<!-- proceeding entry -->
<dl>
<dd>
<div class="meta">
<div class="list-title mathjax"> <span class="descriptor">Title: "A deep learning approach for Magnetic Resonance Fingerprinting"</span> </div>
<div class="list-authors"><span class="descriptor">Authors: M. Golbabaee, D. Chen, P. Gomez, M. Menzel and M. Davies</span> </div>
</div>
</dd>
</dl>
<i>Remark: This iTWIST'18 2-page paper is related to this longer article by
the same authors: <a href="https://arxiv.org/abs/1809.01749">arXiv:1809.01749</a></i>

<hr>
<h3>Proceedings of iTWIST'18, Paper-ID: 35, Marseille, France, November, 21-23, 2018.</h3>
<!-- proceeding entry -->
<!-- <dl> -->
<!-- <dd> -->
<!-- <div class="meta"> -->
<!-- <div class="list-title mathjax"> <span class="descriptor">Title: "Phase inpainting in time-frequency plane"</span> </div> -->
<!-- <div class="list-authors"><span class="descriptor">Authors: A. M. Krémé, V. Emiya and C. Chaux</span> </div> -->
<!-- </div> -->
<!-- </dd> -->
<!-- </dl> -->
LIST:arXiv:1812.04311

<hr>
<h3>Proceedings of iTWIST'18, Paper-ID: 36, Marseille, France, November, 21-23, 2018.</h3>
<!-- proceeding entry -->
<!-- <dl> -->
<!-- <dd> -->
<!-- <div class="meta"> -->
<!-- <div class="list-title mathjax"> <span class="descriptor">Title: "Single molecule localization by L<sub>2</sub>-L<sub>0</sub> constrained optimization"</span> </div> -->
<!-- <div class="list-authors"><span class="descriptor">Authors: Arne Bechensteen, Laure Blanc-Féraud and Gilles Aubert</span> </div> -->
<!-- </div> -->
<!-- </dd> -->
<!-- </dl> -->
LIST:arXiv:1812.05971

<hr>
<h3>Proceedings of iTWIST'18, Paper-ID: 37, Marseille, France, November, 21-23, 2018.</h3>
<!-- proceeding entry -->
<dl>
<dd>
<div class="meta">
<div class="list-title mathjax"> <span class="descriptor">Title: "Bilinear Regression via Convex Programming without Lifting"</span> </div>
<div class="list-authors"><span class="descriptor">Authors: S. Bahmani</span> </div>
</div>
</dd>
</dl>
<i>Remark: This iTWIST'18 2-page paper is related to this longer article by
the same author: <a href="https://arxiv.org/abs/1806.07307">arXiv:1806.07307</a></i>

<hr>
<h3>Proceedings of iTWIST'18, Paper-ID: 38, Marseille, France, November, 21-23, 2018.</h3>
<!-- proceeding entry -->
<dl>
<dd>
<div class="meta">
<div class="list-title mathjax"> <span class="descriptor">Title: "A Low Rank Approach to Off-The-Grid Sparse Super-Resolution"</span> </div>
<div class="list-authors"><span class="descriptor">Authors: P. Catala, V. Duval and G. Peyré</span> </div>
</div>
</dd>
</dl>
<i>Remark: This iTWIST'18 2-page paper is related to this longer article by
the same authors: <a href="https://arxiv.org/abs/1712.08800">arXiv:1712.08800</a></i>
<br />
<hr>
<br />
<br />
<center>
Proceedings of iTWIST'18, Published on <a href="https://arxiv.org/">arXiv.org</a> in December 2018.
</center>

</body>
</html>